\documentclass{ws-mpla}

\begin{document}

\markboth{Li Zhao, Yu-Xiao Liu and Yi-Shi Duan} {Fermions in
gravity and gauge backgrounds on a brane world}

\catchline{}{}{}{}{}


\title{FERMIONS IN GRAVITY AND GAUGE BACKGROUNDS ON A BRANE WORLD}

\author{Li Zhao,
 Yu-Xiao Liu\footnote{Corresponding author. E-mail: liuyx@lzu.edu.cn},
 Yi-Shi Duan}

\address{Institute of Theoretical Physics, Lanzhou University,
   Lanzhou 730000, P. R. China \\
   \vspace{2mm}
Email: zhl03@lzu.cn, liuyx@lzu.edu.cn, ysduan@lzu.edu.cn }

\maketitle


\begin{abstract}
We solve the fermionic zero modes in gravity and gauge backgrounds
on a brane involving a warped geometry, and study the localization
of spin 1/2 fermionic field on the brane world. The result is that
there exist massless spin 1/2 fermions which can be localized on
the bulk with the exponentially decreasing warp factor if
including $U(1)$ gauge background. Two special cases of gauge
backgrounds on the extra dimensional manifold are discussed.

\end{abstract}

\ccode{PACS Nos.: 04.50.+h, 11.25.-w}

\section{Introduction}
It is nowadays widely believed that extra dimensions play an
important role in constructing a unified theory of all
interactions and provides us with a new solution to hierarchy
problem.
\cite{Rubakov1983,Akama,Visser,Antoniadis1990,Dvali,Arkani-Hamed,AADDPLB1998,RS1}
(for a review see, e.g. Ref. \refcite{Lorenzana}). Ref.
\refcite{AADDPLB1998} gave the first string realization of low
scale gravity and braneworld models and pointed out the motivation
of TeV strings from the stabilization of mass hierarchy. Modifying
the old Kaluza-Klein picture, \cite{Bailin} the recent
developments are based on the idea that ordinary matter fields
could be confined to a three-dimensional world, corresponding to
our apparent Universe, while gravity could live in some
higher-dimensional space-time. In the later, Gogberashvili
\cite{GogberashviliIJMPD2002,GogberashviliEPL2000} and Randall and
Sundrum \cite{RS1,RS2} reviving the old idea, \cite{Rubakov1983}
have recently pointed out that the extra dimension need not even
be compact.

Following the brane world models proposed by Randall and Sundrum
(RS), a fair amount of activity has been generated involving
possible extensions and generalizations, among which, co-dimension
two models in six dimensions have been a topic of increasing
interest.\cite{Chodos,CohenPLB1999,Gherghetta,Oda,Ponton,KantiPRD2001}
A useful review on topological defects in higher dimensional
models and its relation to braneworlds is available in Ref.
\refcite{Roessl}.

On the other hand, the other local fields except the gravitational
field are not always localized on the brane even in the warped
geometry. This localization mechanism has been recently
investigated within the framework of a local field theory. It has
been shown that the graviton \cite{RS2} and the massless scalar
field \cite{B.Bajc} have normalizable zero modes on branes of same
types, that the Abelian vector fields are not localized in the RS
model in five dimensions but can be localized in some
higher-dimensional generalizations of it.\cite{I.Oda}  Moreover,
spin 1/2 and 3/2 fermionic fields are localized on a brane with
negative tension.\cite{B.Bajc,Grossman} Thus, in order to fulfill
the localization of fermionic fields on a brane with positive
tention, it seems that some additional interactions except the
gravity should be introduced in the
bulk.\cite{LiuJHEP2007,Liu0708.0065}

Since spin half fields can not be localized on the brane
\cite{RS1,RS2,Oda} in five or six dimensions by gravitational
interaction only, it becomes necessary to introduce additional
non-gravitational interactions to get spinor fields confined to
the brane or string-like defect. The mechanism of localization of
spin 1/2 fermions on a brane was first discovered in a flat
space-time long ago by Jackiw and Rebbi \cite{Jackiw1976} and
recently extended to the case of $AdS_5$ by Grossman and
Neubert.\cite{Grossman} More recently, Randjbar-Daemi $\it et\; al
$ studied localization of bulk fermions on a brane with inclusion
of scalar backgrounds \cite{Randjbar-Daemi} and minimal gauged
supergravity \cite{Parameswaran} in higher dimensions and gave the
conditions under which localized chiral fermions can be obtained.
Motivated by the inclusion of the bulk
scalars,\cite{Randjbar-Daemi} in this letter, we carry out our
search for the  gauge fields on a 3-brane in six space-time
dimensions instead of the real scalar field  on this issue. It is
shown that spin 1/2 spinor field is confined on the $3$-brane
without appealing to the additional bulk interactions except the
gravity and gauge fields.

This paper is organized as follows: In Sec. \ref{secDiracEq}, we
obtain the effective Lagrangian of  fermions in gravity and gauge
backgrounds. In Sec. \ref{gaugeFields}, we give two ansatz of the
background $U(1)$ gauge fields and the conditions under which the
gauge fields satisfy the equation of motion for the vector fields.
In Sec. \ref{seclocalization}, we solve the fermionic zero modes and
check the localization of bulk fermions on a $3$-brane under two
simple assumptions for the $U(1)$ gauge fields. In the last section,
a brief conclusion is presented.


\section{Dirac equation in gravity and gauge  backgrounds}\label{secDiracEq}
We shall consider the six-dimensional generalizations of the RS
model with the warped geometry
(the $(+,-,-,-,-,-)$ signature will be assumed below):
\begin{equation} \label{Metric}
ds^2=e^{A(r)}\eta_{\mu \nu} dx^\mu dx^\nu
-dr^{2}-e^{B(r)}a^{2}d\theta^{2},
\end{equation}
where   $\eta_{\mu\nu}$ is the ordinary flat Minkowski metric, $a$
is the radius of the
circle covered by the coordinate $\theta$. 
For the two extra spatial dimensions we have introduced polar
coordinates $(r,\theta)$ with $0\leq r<\infty$ and  $0\leq \theta<2
\pi$.

Let us consider the action of a massless spin 1/2 fermions coupled
to gravity
\begin{equation} S=\int d^{6}x \sqrt{-g}{\bar
\Psi}\Gamma^{A}E^{M}_{A}D_{M}\Psi,
\end{equation}
where $A=0,\cdots,5$ corresponds to the flat tangent six-dimensional
Minkowski space, $M=0,\cdots, 5$ denotes the six-dimensional
spacetime index. The corresponding equation of motion takes the form
\begin{equation} \label{DiracEq1}
\Gamma^A E^{M}_{A} (\partial_M + \Omega_M  +  A_M)
\Psi(x,r,\theta)=0,
\end{equation}
where $E^{A}_{M}$ is the {\sl sechsbein} with $E^{A}_{M} =
(e^{\frac{A(r)}{2}}\delta^{a}_{ \mu},\;1,\;ae^{\frac{B(r)}{2}}),$
$\Omega_M =\frac{1}{4}\Omega_{M}^{AB}\Gamma_{A}\Gamma_{B}$ is the
spin connection and $A_{M}$ is $U(1)$ gauge fields. The RS model
is the special case with $M=0,\cdots, 4$ and $A_M = 0$. The spin
connection $\Omega_M^{AB}$ is defined as
\begin{eqnarray}
\Omega_M^{AB}
 &=& \frac{1}{2}E^{NA}(\partial_{M}E^{B}_{N}-\partial_{N}E^{B}_{M}) 
 - \frac{1}{2}E^{NB}(\partial_{M}E^{A}_{N}-\partial_{N}E^{A}_{M}) \nonumber \\
 &-& \frac{1}{2}E^{PA}E^{QB}
            (\partial_{P}E_{QC}-\partial_{Q}E_{PC})E^{C}_{M}.
 \nonumber
\end{eqnarray}
So the non-vanishing components of $\Omega_M$ are
\begin{equation}
\Omega_{\mu}=\frac{1}{4}e^{\frac{A(r)}{2}}A^{'}(r)\delta^{a}_{\mu}\Gamma_{a}\Gamma_{4},\quad
\Omega_{5}=\frac{1}{4}ae^\frac{B(r)}{2}B^{'}(r)\Gamma_{5}\Gamma_{4},
\end{equation}
where the prime denotes the derivative with respect to $r$.
Substituting the non-vanishing components of $\Omega_M$ into the
six-dimensional Dirac equation (\ref{DiracEq1}) gives
\begin{equation} \label{DiracEq2}
\left \{e^{-\frac{A}{2}}\Gamma^a \delta^{\mu}_{a} (\partial_{\mu} +
A_{\mu}) + \Gamma^4 \left(\partial_{r} +
A_{r}+A^{'}(r)+\frac{B^{'}(r)}{4}\right)+
\frac{1}{a}\Gamma^5e^{-\frac{B}{2}}(\partial_{\theta}+A_{\theta})
\right \} \Psi = 0. \nonumber
\end{equation}
We denote the Dirac operators on the four-dimensional manifold $M$
and the two extra dimensional manifold $K $ with $D_M$ and $D_K$,
respectively:
\begin{eqnarray}
D_M &=& e^{-\frac{A}{2}}\bar{\Gamma} \Gamma^a \delta^{\mu}_{a}
(\partial_{\mu}
+  A_{\mu}),\\
D_K &=& \bar{\Gamma}\left \{\Gamma^4 \left(\partial_{r} +
A_{r}+A^{'}(r)+\frac{B^{'}(r)}{4}\right)+
\frac{1}{a}\Gamma^5e^{-\frac{B}{2}}(\partial_{\theta}+A_{\theta})
\right \},
\end{eqnarray}
where $\bar{\Gamma}=\Gamma^0 \Gamma ^1 \Gamma ^2\Gamma ^3 $ and
$D_K$ is a kind of `mass' operator whose operator eigenvalues are
fermion masses as seen in four dimensions. Then we have the
following commutative relations:
\begin{equation}
[D_M,D_K]=0, \nonumber
\end{equation}
and can expand any spinor $\Psi$ in a set of eigenvectors $\phi_{m}$
of the operator $D_K$
\begin{equation}
D_K\phi_{m}=\lambda_{m}\phi_{m}.
\end{equation}
Thus, each $\phi_m$ is observed in four dimensions as a fermion of
mass $\lambda_{m}$. All these eigenvalues play a role of the mass
of the corresponding four-dimensional excitations.\cite{Libanov}
We assume that the energy scales probed by a four-dimensional
observer are smaller than the separation, and thus even the first
non-zero level is not excited. This implies that we are  looking
for the solutions  of the zero modes of $D_K$
\begin{equation} \label{DiracEqOnT2}
D_K\phi=0.
\end{equation}
It is just the Dirac equation on the manifold $K$. For fermionic
zero modes, we have the following decomposition
\begin{equation}
\Psi(x,r,\theta)=\psi(x) \phi(r,\theta),
\end{equation}
where  $\phi(r,\theta)$ satisfies Eq. (\ref{DiracEqOnT2}). The
effective Lagrangian for $\psi(x)$ is defined as
\begin{eqnarray}
\mathcal{L}_{eff}&=&\int dr d\theta  \sqrt{-g}
\bar{\Psi} \Gamma^A E^{M}_{A} (\partial_M + \Omega_M +  A_M) \Psi  \nonumber \\
&=& \bar{\psi} \Gamma^a \delta^{\mu}_{a} (\partial_{\mu}  +
A_{\mu}) \psi(x) \int  dr d\theta e^{-\frac{A(r)}{2}}\sqrt{-g}
\phi^{\dag}\phi\nonumber \\
&=& a\bar{\psi} \Gamma^a \delta^{\mu}_{a} (\partial_{\mu}  +
A_{\mu}) \psi(x)  \int  dr d\theta e^{\frac{3A(r)+B(r)}{2}}
\phi^{\dag}\phi.
\end{eqnarray}
Thus, to have the localization of finite kinetic energy for
$\psi(x)$, the above integral must be finite. This can be achieved
if the function $\phi(r,\theta)$ does not diverge on the extra
dimensional manifold $K$.

\section{Equation of motion for the $U(1)$ gauge fields}
\label{gaugeFields}

Let us turn to the $U(1)$ gauge fields. Here we consider the
action of the spin $1$ vector fields
\begin{eqnarray}
S_1 = - \frac{1}{4} \int d^5 x \sqrt{-g} g^{M N} g^{R S} F_{MR}
F_{NS}, \label{44}
\end{eqnarray}
where the gauge field tensor $F_{MN} = \partial_M A_N - \partial_N
A_M$ as usual with $A_{M}$ the gauge fields.  In this letter, to
simplify the analysis and without loss of generality, it is
assumed that the gauge fields $A_{\mu}$, $A_{r}$ and $A_{\theta}$
satisfy the following two ansatz: \textbf{Ansatz I}:
$A_{\mu}=A_{\mu}(x),A_{r}=A_{r}(r),A_{\theta}=A_{\theta}(\theta)$
and \textbf{Ansatz II}: $A_{\mu}=A_{\mu}(x),
A_{r}=A_{r}(r),A_{\theta}=A_{\theta}(r)$, which are the function
of the four dimensional spacetime coordinate $x$ and extra
dimensional spacetime coordinate $r$ and $\theta$, respectively.
Then one may doubt whether the two different ansatz of gauge
fields $A_{r}$ and $A_{\theta}$ satisfy the equation of motion for
the vector fields. This property will be reflected mathematically
in the following. From the action (\ref{44}), the equation of
motion is given by
\begin{eqnarray}
\frac{1}{\sqrt{-g}} \partial_M (\sqrt{-g} g^{M N} g^{R S} F_{NS}) =
0, \label{eom}
\end{eqnarray}
which can be expanded as
\begin{equation}
\frac{1}{\sqrt{-g}} \partial_\mu(\sqrt{-g} g^{\mu\nu} g^{RS} F_{\nu
S})+\frac{1}{\sqrt{-g}} \partial_i (\sqrt{-g} g^{ij} g^{R S} F_{jS})
= 0.\label{equationofmotion}
\end{equation}
Note that in this letter $\mu,\nu,\lambda$ denote the four
dimensional spacetime indices, and $i,j,k$ denote the extra
dimensional spacetime indices. To analyze this equation in more
detail, we divide the index $R$  into the following three cases:

\textbf{Case I: $R=\tau$}

From the background geometry (\ref{Metric}), Eq.
(\ref{equationofmotion}) changes into
\begin{equation}
\frac{1}{\sqrt{-g}} \partial_\mu(\sqrt{-g} g^{\mu\nu}
g^{\tau\lambda} F_{\nu \lambda})+\frac{1}{\sqrt{-g}} \partial_i
(\sqrt{-g} g^{ij} g^{\tau\lambda} F_{j\lambda}) = 0,
\label{equationofmotion'}
\end{equation}
where the first term on the LHS of (\ref{equationofmotion'}) is the
usual  equation of motion in four dimensional spacetime. Due to the
assumption of $A_{j}=A_{j}(r,\theta), A_{\lambda}=A_{\lambda}(x)$,
the gauge field tensor $
F_{j\lambda}=\partial_{j}A_{\lambda}(r,\theta)-\partial_{\lambda}A_{j}(x)
$ naturally equals to zero. Thus, under the requirement that the
$A_{\mu}$ meets the usual equation of motion in four dimensional
spacetime, the two different ansatz of gauge fields $A_{r}$ and
$A_{\theta}$ satisfy the equation of motion for the vector fields in
six dimensional spacetime.

\textbf{Case II:  $R=r$}

One can find that the  equation (\ref{eom}) changes to
\begin{equation}
\frac{1}{\sqrt{-g}} \partial_\mu(\sqrt{-g} g^{\mu\nu} g^{rr} F_{\nu
r})+\frac{1}{\sqrt{-g}} \partial_i (\sqrt{-g} g^{ij} g^{rr} F_{jr})
= 0. \label{equationofmotion1}
\end{equation}
It is easy to see that $F_{\nu r}$  equals to zero, so we only
calculate the value of second term on the LHS of
(\ref{equationofmotion1}). Considering that  $F_{jr}$ is an anti
symmetric tensor, then $F_{rr}=0$,  so we only need to consider
the case of $j=\theta$. For the first ansatz
$A_{r}=A_{r}(r),A_{\theta}=A_{\theta}(\theta)$, the gauge field
tensor is
\begin{equation}
F_{\theta r}=\partial_{\theta}A_{r}(r)-\partial_{r}A_{\theta}(\theta),
\end{equation}
which also equals to zero. This situation is similar to case I.
However, for the second ansatz
$A_{r}=A_{r}(r),A_{\theta}=A_{\theta}(r)$, we have
\begin{eqnarray}
 F_{\theta r}&=&
      \partial_{\theta}A_{r}(r)
      -\partial_{r}A_{\theta}(r) \nonumber \\
    &=& -\partial_{r}A_{\theta}(r).
\end{eqnarray}
Then equation (\ref{equationofmotion1}) is reduced to
\begin{eqnarray}
 \partial_{\theta}(\sqrt{-g}
 g^{\theta\theta}g^{rr}\partial_{r}A_{\theta}(r))=0,
 \label{eom'}
\end{eqnarray}
which is true for any $A_{r}(r)$ and $A_{\theta}(r)$. So in the
case of $R=r$, both ansatz are allowed.

\textbf{Case III: $R=\theta$}

Equation (\ref{eom}) becomes
\begin{equation}
\partial_\mu(\sqrt{-g} g^{\mu\nu}
g^{\theta\theta} F_{\nu \theta})+ \partial_r (\sqrt{-g} g^{rr}
g^{\theta\theta} F_{r\theta}) = 0,
\end{equation}
which is reduced to
\begin{equation}
\partial_r (\sqrt{-g} g^{rr} g^{\theta\theta}
F_{r\theta}) = 0 \label{equationofmotion2}
\end{equation}
for $F_{\nu \theta}=0$. For the first ansatz $A_{r}=A_{r}(r)$ and
$A_{\theta}=A_{\theta}(\theta)$, the gauge field tensor $F_{
r\theta}=\partial_{r}A_{\theta}-\partial_{\theta}A_{r}=0$. While
for the second ansatz $A_{r}=A_{r}(r)$ and
$A_{\theta}=A_{\theta}(r)$, one has
$F_{r\theta}=\partial_{r}A_{\theta}(r)$. Then the equation of
motion reduces to
\begin{equation}
\partial_{r} (e^{2A(r)-\frac{B(r)}{2}}\partial_{r}A_{\theta}(r))=0.\label{equationofmotion3}
\end{equation}
It suggests that, in order to satisfy the equation of motion,
there exists a constraint (\ref{equationofmotion3}) for
$A_{\theta}(r)$. In fact, this equation can be simply satisfied.
For example, a simple choice is $A_{\theta}(r)=C$ with $C$ a
constant.

From the above analysis, it can be concluded that, in order to
satisfy the equation of motion for the gauge fields,  there should
exist one constraint equation (\ref{equationofmotion3})  for the
gauge fields $A_{\theta}(r)$ under the second ansatz. While for
the first ansatz, one need not any constraint.

\section{Fermionic zero modes and localization of fermions}
\label{seclocalization}

In this section, we solve the fermionic zero modes under the above
two ansatz for the gauge fields and discuss the localization of
the Dirac fermions in these gauge backgrounds. We have the
physical setup in mind such that ``local cosmic string" sits at
the origin $r = 0$ and then ask the question of whether various
bulk fermions with spin 1/2  can be localized on the brane with
the exponentially decreasing warp factor by means of the
gravitational interaction and gauge background. Of course, in due
analysis, we will neglect the backreaction on the geometry induced
by the existence of the bulk fields.

\subsection{Ansatz I:  $A_{\mu}=A_{\mu}(x),
A_{r}=A_{r}(r)$ and $A_{\theta}=A_{\theta}(\theta)$}

For our current ansatz, the Dirac equation (\ref{DiracEqOnT2}) is
read as
\begin{equation} \label{DiracEqCaseI}
\bar{\Gamma}\left \{ \Gamma^4
\left(\partial_{r}+A_{r}(r)+A^{'}(r)+\frac{B^{'}(r)}{4}
\right)+\frac{1}{a}\Gamma^5
e^{-\frac{B}{2}}(\partial_{\theta}+A_{\theta}(\theta))\right \} \phi
= 0.
\end{equation}
We are now ready to study the above Dirac equation for
 $6$-dimensional fluctuations, and write it in terms of
$4$-dimensional effective fields. Since $\phi$ is a $6$-dimensional
Weyl spinor we can represent it by
\begin{equation}
\phi(r,\theta)=
\left(%
\begin{array}{c}
  \phi_{1}^{(4)} \\
  \phi_{2}^{(4)} \\
\end{array}%
\right),
\end{equation}
where $\phi_{1}^{(4)}$ and  $ \phi_{2}^{(4)}$ are $4$-dimensional
Dirac spinors. Our choice for the 6-dimensional constant gamma
matrices $\Gamma^A, A = 0, 1, 2, 3, 4,5$ ¯µ are:
\begin{equation}
\Gamma^A=\begin{pmatrix}
  0 & \Sigma^A \\
  \bar{\Sigma}^A & 0 \\
\end{pmatrix}. \nonumber
\end{equation}
Here $\Sigma^0 = \bar{\Sigma}^0 = \gamma^0 \gamma^0$; $\Sigma^i =
-\bar{\Sigma}^i = \gamma^0 \gamma^i$; $\Sigma^4 = -\bar{\Sigma}^4 =
i\gamma^0 \gamma^5$; $\Sigma^5 = -\bar{\Sigma}^5 = \gamma^0$,
$\gamma^{\mu}$ and $\gamma^5$ are usual four-dimensional Dirac
matrices in the chiral representation:
\begin{equation}
\gamma^0=\begin{pmatrix}
  0 & 1 \\
  1 & 0 \\
\end{pmatrix},\;\;\;
\gamma^i=\begin{pmatrix}
  0 & \sigma^i \\
  -\sigma^i & 0 \\
\end{pmatrix},\;\;\;
\gamma^5=i \gamma^0\gamma^1\gamma^2\gamma^3=\begin{pmatrix}
  1 & 0 \\
  0 & -1 \\
\end{pmatrix} \nonumber
\end{equation}
with $\sigma^i$ the Pauli matrices. $\Gamma^A$ have the relation
$\Gamma^A\Gamma^B+\Gamma^B\Gamma^A=2\eta^{AB}I$. Then the Dirac
equation (\ref{DiracEqCaseI}) can be reduced to
\begin{eqnarray}
\left \{ae^{\frac{B}{2}}
\left(%
\begin{array}{cc}
  0 & \gamma^{0} \\
  \gamma^{0} & 0 \\
\end{array}%
\right)\left(\partial_{r}+A_{r}(r)+A^{'}(r)
+\frac{B^{'}(r)}{4}\right) \right. \nonumber \\
+ \left. \left(%
\begin{array}{cc}
  0 & i\gamma^{5} \gamma^{0}\\
  i\gamma^{5} \gamma^{0} & 0 \\
\end{array}%
\right) (\partial_{\theta}+A_{\theta}(\theta))\right \} \left(
\begin{array}{c}
  \phi_{1}^{(4)} \\
  \phi_{2}^{(4)} \\
\end{array}
\right)
 =0.
\end{eqnarray}
Obviously, the solutions of $ \phi_{1}^{(4)}$ and $\phi_{2}^{(4)}$
are the same. For simplicity and without loss of generality, we
consider the solution of  $\phi_{1}^{(4)}$. Denoting
\begin{equation}
\phi_{1}^{(4)}=\left(
\begin{array}{c}
  \phi_{11}^{(2)} \\
   \phi_{12}^{(2)} \\
  \end{array}
\right)
\end{equation}
with $\phi_{11}^{(2)}$ and $\phi_{12}^{(2)}$ the $2$-dimensional
Dirac spinors, one can obtain the following differential equations
\begin{equation}
\left
\{ae^{\frac{B}{2}}\sigma^{0}\left(\partial_{r}+A_{r}(r)+A^{'}(r)+\frac{B^{'}(r)}{4}\right)+
iI_{2\times2}(\partial_{\theta}+A_{\theta}(\theta))\right \}
  \phi_{11}^{(2)} \\
=0,
\end{equation}
\begin{equation}
\left
\{ae^{\frac{B}{2}}\sigma^{0}\left(\partial_{r}+A_{r}(r)+A^{'}(r)+\frac{B^{'}(r)}{4}\right)-
iI_{2\times2}(\partial_{\theta}+A_{\theta}(\theta))\right \}
  \phi_{12}^{(2)}
=0.
\end{equation}
Solving  the  above two equations one can easily get the formalized
solutions:
\begin{equation}
\left\{
\begin{array}{l}\label{decomposed solution}
\phi_{11}^{(2)}(r,\theta) =e^{-A(r)-\frac{B(r)}{4}-\int dr A_{r}(r)
+ \frac{iC}{a}\int dr e^{-\frac{B(r)}{2}}+C\theta-\int d\theta
A_{\theta}(\theta)}, \\
\phi_{12}^{(2)}(r,\theta) = e^{-A(r)-\frac{B(r)}{4}-\int dr A_{r}(r)
-\frac{iC}{a}\int dr e^{-\frac{B(r)}{2}}+C\theta-\int d\theta
A_{\theta}(\theta)}
\end{array}
\right.
\end{equation}
with $C$ being an integration constant. Because of the  {\sl
sechsbein} transformation properties, $\phi_{i}^{(4)}(r,\theta)$
has to be antiperiodic,\cite{Randjbar-Daemi2003}
$\phi_{i}^{(4)}(r,\theta)=-\phi_{i}^{(4)}(r,\theta+2\pi)$, then we
get $C=i(n+\frac{1}{2})\;(n\in Z). $ Therefore, in the case of the
gauge fields $A_{r}=A_{r}(r)$ and $A_{\theta}=A_{\theta}(\theta)$,
by substituting the constant $C$ into the above expression
(\ref{decomposed solution}), one can get the fermionic zero modes
$\phi(r,\theta)$
\begin{equation}
\phi(r,\theta)= \left(
\begin{array}{c}
  1 \\
  1 \\
 \end{array}
\right) \otimes \left(
\begin{array}{c}
  e^ {-i(n+\frac{1}{2})\theta-A(r)-\frac{B(r)}{4}-\frac{(n+\frac{1}{2})}{a}\int dr e^{-\frac{B(r)}{2}}
 -\int dr  A_{r}(r) -\int d\theta A_{\theta}(\theta)  }\\
  e^{-i(n+\frac{1}{2})\theta-A(r)-\frac{B(r)}{4}+\frac{(n+\frac{1}{2})}{a}\int dr e^{-\frac{B(r)}{2}}
- \int dr  A_{r}(r) -\int d\theta A_{\theta}(\theta)} \\
\end{array}
\right) \otimes \left(
\begin{array}{c}
  1 \\
  1 \\
 \end{array}
\right). \nonumber
\end{equation}
 The effective Lagrangian for $\psi(x)$ then becomes
\begin{eqnarray}
\mathcal{L}_{eff}&=& \int d\theta d\varphi \sqrt{-g}
\bar{\Psi} \Gamma^A E^{M}_{A} (\partial_M - \Omega_M +  A_M) \Psi \nonumber\\
&=& 4a\bar{\psi} \Gamma^a \delta^{\mu}_{a}
\partial_{\mu} \psi I_{1}I_{2},
\label{Lagrangian}
\end{eqnarray}
where
\begin{equation}
I_{1}=2\int_{0} ^{\infty} dr e^{-\frac{1}{2}A(r)-2\int dr A_{r}(r)
} \cosh\left ( \frac{2n+1}{a}\int dr e^{-\frac{B(r)}{2}} \right),
\label{I1}
\end{equation}
 and
\begin{equation}I_{2}= \int_{0}
^{2\pi} d\theta e^{-2\int{d \theta A_{\theta}(\theta)}}. \label{I2}
\end{equation}
In order to localize spin 1/2 fermions in this framework, the
integrals (\ref{I1}) and (\ref{I2}) should be finite. By
considering the Einstein's equation without sources, the solutions
of the metric functions $A(r)$ and $B(r)$ are given by
\cite{Gregory}
\begin{equation}
A(r)=B(r)=-cr, \label{metrical function}
\end{equation}
where the parameter $c$ is the combination of the bulk
cosmological constant and the Newton constant. Therefore, for such
exponential warp factors $A(r)$ and $B(r)$, to have localized
fermions, it is sufficient if $I_{1}=\int_{0}^{\infty} dr
e^{\frac{1}{2}cr-2\int dr A_{r}(r) } \cosh\left ( \frac{4n+2}{a
c}e^{\frac{cr}{2}}\right)$ and $I_{2}=\int_{0} ^{2\pi} d\theta
e^{-2\int{d \theta A_{\theta}(\theta)}}$ are finite on $K$. When
the gauge background vanishes, this integral is obviously
divergent for the exponentially decreasing warp factor $c > 0$
while it is finite for the exponentially increasing warp factor $c
< 0$. This situation is the same as in the case of the domain wall
in the RS framework \cite{Bajc} where for localization of spin
$1/2 $ field additional localization method by Jackiw and Rebbi
\cite{Jackiw1976} was introduced.  However, if the gauge
backgrounds are considered,  the integrals $I_{1}$ and $I_{2}$ can
be normalizable for not only the exponentially increasing but also
the exponentially decreasing warp factor.

\subsection{ Ansatz II: $A_{\mu}=A_{\mu}(x), A_{r}=A_{r}(r)$ and $A_{\theta}=A_{\theta}(r)$}

In this simple assumption, note that $A_{\theta}(r)$ should be
constrained by equation (\ref{equationofmotion3}) while $A_{r}(r)$
is an arbitrary function. But we shall prove that $A_{\theta}(r)$
have no contribution to the effective Lagrangian. The Dirac
equation (\ref{DiracEqOnT2}) becomes
\begin{equation}
 \bar{\Gamma}\left \{ \Gamma^4
  \left(\partial_{r}+A^{'}(r)+\frac{B^{'}(r)}{4}
        +A_{r}(r)+\Gamma^{5}\Gamma^{4}a^{-1}e^{-\frac{B}{2}}A_{\theta}(r)
  \right)
  +\frac{1}{a}\Gamma^5 e^{-\frac{B}{2}}\partial_{\theta}\right \} \phi = 0.
  \label{DiracEqCaseI1}
\end{equation}
Denoting
\begin{equation}
\phi(r,\theta)=
\left(%
\begin{array}{c}
  \phi_{1}^{(4)} \\
  \phi_{2}^{(4)} \\
\end{array}%
\right) ,
\end{equation}
the Dirac equation (\ref{DiracEqCaseI1}) reduces to
\begin{eqnarray}
\left \{ae^{\frac{B}{2}}
\left(%
\begin{array}{cc}
  0 & \gamma^{0} \\
  \gamma^{0} & 0 \\
\end{array}%
\right)
\left(\partial_{r}+A^{'}(r)+\frac{B^{'}(r)}{4}+A_{r}(r)+ia^{-1}e^{-\frac{B}{2}}
\gamma^{5}\otimes I_{2\times2}A_{\theta}(r)\right) \right.  \nonumber \\
 + \left. \left(%
\begin{array}{cc}
  0 & i\gamma^{5} \gamma^{0}\\
  i\gamma^{5} \gamma^{0} & 0 \\
\end{array}%
\right) \partial_{\theta}\right \} \left(
\begin{array}{c}
  \phi_{1}^{(4)} \\
  \phi_{2}^{(4)} \\
\end{array}
\right) =0,\nonumber
\end{eqnarray}
i.e.
\begin{equation}
\left \{ae^{\frac{B}{2}}
\gamma^{0}\left(\partial_{r}+A^{'}(r)+\frac{B^{'}(r)}{4}+A_{r}(r)+ia^{-1}e^{-\frac{B}{2}}
\gamma^{5}A_{\theta}(r)\right)+ i\gamma^{5}
\gamma^{0}\partial_{\theta}\right \}
  \phi_{i}^{(4)}
 =0,
 \nonumber
\end{equation}
where i=1,2. Now we come to the issue of the chirality of
fermions, the eight-component spinor $\phi_{i}$ can be written in
terms of the left and right handed spinors
\begin{equation}
\phi_{i}^{(4)}=\phi_{i} ^{(4)L}+\phi_{i}^{(4)R},
\end{equation}
and one gets the system of two equations for the chiral
components. For the left spinors $\gamma^5
\phi_{i}^{(4)L}(r,\theta)=-\phi_{i}^{(4)L}(r,\theta)$ we have
\begin{equation}
\left \{ae^{\frac{B}{2}}
\gamma^{0}\left(\partial_{r}+A^{'}(r)+\frac{B^{'}(r)}{4}+A_{r}(r)
-ia^{-1}e^{-\frac{B}{2}}A_{\theta}(r) \right)+
i\gamma^{0}\partial_{\theta}\right \}
  \phi_{i}^{(4)L}
 =0,
 \nonumber
\end{equation}
and, for the right spinors $\gamma^5
\phi_{i}^{(4)R}(r,\theta)=\phi_{i}^{(4)R}(r,\theta)$
correspondingly
\begin{equation}
\left \{ae^{\frac{B}{2}}
\gamma^{0}\left(\partial_{r}+A^{'}(r)+\frac{B^{'}(r)}{4}+A_{r}(r)
+ia^{-1}e^{-\frac{B}{2}}A_{\theta}(r)\right )-
i\gamma^{0}\partial_{\theta}\right \}
  \phi_{i}^{(4)R}
 =0,
  \nonumber
\end{equation}
By solving the  above  differential equation similarly as the above
case, one can get the solutions
\begin{eqnarray}
\phi_{i}^{(4)L}(r,\theta)&=&e^ {\left\{-A(r)-\frac{B(r)}{4}-\int
dr
\left(A_{r}(r)-Ca^{-1}e^{-\frac{B(r)}{2}}-ia^{-1}e^{-\frac{B(r)}{2}}A_{\theta}(r)\right)
+iC\theta \right\}}
\left(%
\begin{array}{cc}
  0\\
  0\\
  1\\
  1\\
\end{array}%
\right),\nonumber \\
\phi_{i}^{(4)R}(r,\theta)&=&e^ {\left\{-A(r)-\frac{B(r)}{4}-\int
dr
\left(A_{r}(r)-Ca^{-1}e^{-\frac{B(r)}{2}}+ia^{-1}e^{-\frac{B(r)}{2}}A_{\theta}(r)\right)
-iC\theta \right\}}
\left(%
\begin{array}{cc}
  1\\
  1\\
  0\\
  0\\
\end{array}%
\right),
 \nonumber
\end{eqnarray}
with $C$ an integration constant. To guarantee   the
antiperiodicity of $\phi^{(4)}_{i}(r,\theta)$, it is easy to get
$C=n+\frac{1}{2}\; (n\in Z). $ Then substituting  $C$  into the
above equations, the zero mode $\phi^{(4)}_i(r,\theta)$ on the two
extra dimensions takes the following form
\begin{eqnarray}
\phi^{(4)}_i&=&\phi^{(4)L}_i+\phi^{(4)R}_i\nonumber\\ &=& \left(
\begin{array}{c}
e^ {\left \{-i(n+\frac{1}{2})\theta-A(r)-\frac{B(r)}{4}-\int  dr
\left(A_{r}(r)-\frac{2n+1}{2a}e^{-\frac{B(r)}{2}}+ia^{-1}e^{-\frac{B}{2}}A_{\theta}(r)\right)
  \right \}}\\
e^{\left \{+i(n+\frac{1}{2})\theta-A(r)-\frac{B(r)}{4}-\int dr
\left(A_{r}(r)-\frac{2n+1}{2a}e^{-\frac{B(r)}{2}}-ia^{-1}e^{-\frac{B}{2}}A_{\theta}(r)\right)
\right \}} \\
\end{array}
\right) \otimes \left(
\begin{array}{c}
  1 \\
  1 \\
 \end{array}
\right). ~~~~~ \nonumber
\end{eqnarray}
The effective Lagrangian for $\psi(x)$ then becomes
\begin{eqnarray}
\mathcal{L}_{eff}&=&\int d\theta d\varphi \sqrt{-g}
\bar{\Psi} \Gamma^A E^{M}_{A} (\partial_M - \Omega_M + A_M) \Psi \nonumber\\
&=& 8\pi a\bar{\psi} \Gamma^a \delta^{\mu}_{a}
\partial_{\mu} \psi I_{3} ,
\end{eqnarray}
where
\begin{equation}
 I_{3}=\int_{0} ^{\infty} dr e^{-\frac{1}{2}A(r)-2\int dr A_{r}(r) +\frac{2n+1}{a}\int dr
e^{-\frac{B(r)}{2}}}.
\end{equation}
This result shows that, whatever the form of $A_\theta(r)$ is, the
effective Lagrangian for $\psi(x)$ has the same form, i.e.,
$A_\theta(r)$ does not affect the effective Lagrangian.
By taking the metric functions $A(r)$ and $B(r)$ in (\ref{metrical
function}), these fermionic zero modes are generically normalizable
on the brane with the use of the gauge fields if the integral
$I_{3}=\int_{0}^{\infty} dr e^{\frac{cr}{2}-2\int dr A_{r}(r)
-\frac{4n+2}{a c} e^{-\frac{cr}{2}}}$ does not diverge, and we need
not include any other bulk field to localize the bulk fermions.

\section{Conclusions}
In conclusion, we have studied two issues, those are, finding the
solutions of fermionic zero modes with two  extra dimensions and
investigating the possibility of localizing the spin 1/2 fermionic
fields on a brane with the exponentially decreasing warp factor.
Localizing the fermionic fields on the brane requires us to
introduce other interactions but gravity. In this letter, we
include the $U(1)$ gauge fields to  study localization of spin 1/2
fermions on a $3$-brane in six-dimensional spacetime. Two special
ansatz of the gauge fields are presented, and the conditions under
which the gauge fields satisfy the equation of motion are
obtained. It is worthwhile to stress that, in the case of
$A_{\mu}(x), A_{r}=A_{r}(r)$ and $A_{\theta}=A_{\theta}(r)$, we
obtain the zero modes for chiral fermions and the effective
Lagrangian for $\psi(x)$ has the same form whatever the gauge
field $A_{\theta}(r)$ is. It is shown that, the effective
Lagrangian for $\psi(x)$ is definitely finite under some
assumption of the gauge fields  in the extra dimensional manifold,
which means that these fermionic zero modes are generically
normalizable. And the localization of the bulk fermions on a brane
with the exponentially decreasing warp factor is achieved if gauge
and gravitational backgrounds are considered.

Moreover, to localize the fermions on the brane or the string-like
defect, there are some other backgrounds could be considered
besides gauge fields and gravity, for example, vortex
background.\cite{WangMPLA2005,DuanMPLA2006} The localization of
the topological Abelian Higgs vortex coupled to fermion can be
fund in our another work \refcite{LiuNPB2007,LiuCTP2007}.

\section*{Acknowledgments}

It is a pleasure to thank Dr. Zhenhua Zhao  for many useful
discussions. This work was supported by the National Natural
Science Foundation of the People's Republic of China (No. 10475034
and No. 10705013), the Doctor Education Fund of Educational
Department of the People's Republic of China (No. 20070730055) and
the Fundamental Research Fund for Physics and Mathematics of
Lanzhou University (No. Lzu07002).


\begin{thebibliography}{0}
\bibitem{Rubakov1983} V.A. Rubakov and M.E. Shaposhnikov, {\it Phys. Lett. B}  {\bf 125},
136 (1983).
\bibitem{Akama}
 K. Akama,
 {\it Lect. Notes Phys.}  {\bf 176},  267 (1982).
\bibitem{Visser}
 M. Visser,
 {\it Phys. Lett. }  {\bf B 159},  22 (1985).

\bibitem{Antoniadis1990}
 I. Antoniadis,
 {\it Phys. Lett.} {\bf B 246}, 317 (1990).

\bibitem{Dvali}
 G.R. Dvali and M.A. Shifman,
 {\it Phys. Lett. }  {\bf B 396}, 64 (1997).
\bibitem{Arkani-Hamed}
 N. Arkani-Hamed, S. Dimopoulos and G.R. Dvali,
 {\it Phys. Lett. }  {\bf B 429}, 263  (1998).

\bibitem{AADDPLB1998}
 I. Antoniadis, N. Arkani-Hamed, S. Dimopoulos and G. Dvali,
 {\it Phys. Lett.} {\bf B 436} (1998) 257.

\bibitem{RS1}
 L. Randall and R. Sundrum,
 {\it Phys. Rev. Lett.} {\bf 83}, 3370 (1999).
\bibitem{Lorenzana}
 A. P\'{e}rez-Lorenzana,
 {\em An Introduction to the Brane World}, arXiv:hep-ph/0406279.

\bibitem{Bailin}
 D. Bailin and A. Love,
 {\it Rep. Prog. Phys.} {\bf 50},  1087 (1987).

\bibitem{GogberashviliIJMPD2002}
 M. Gogberashvili,
 {\it Int. J. Mod. Phys.} {\bf D 11}, 1635 (2002).

\bibitem{GogberashviliEPL2000}
 M. Gogberashvili,
 {\it Europhys. Lett.}  {\bf 49}, 396 (2000).

\bibitem{RS2}
 L. Randall and R. Sundrum,
 {\it Phys. Rev. Lett.} {\bf 83}, 4690 (1999).



\bibitem{Chodos}
 A. Chodos and E. Poppitz,
    {\it Phys. Lett. }   {\bf B 471}, 119 (1999).

\bibitem{CohenPLB1999}
 A.G. Cohen and D.B. Kaplan,
     {\it Phys. Lett. }  {\bf B 470},  52  (1999).

\bibitem{Gherghetta}
 T. Gherghetta and M.E. Shaposhnikov,
 {\it Phys. Rev. Lett.} {\bf 85}, 240 (2000).

\bibitem{Oda}
 I. Oda,
 {\it Phys. Lett. }  {\bf B 496}, 113 (2000).

\bibitem{Ponton}
 E. Ponton and E. Poppitz,
    {\it JHEP} {\bf 02}, 042 (2001).

\bibitem{KantiPRD2001}
 P. Kanti, R. Madden and K.A. Olive,
    {\it Phys. Rev. } {\bf D 64}, 044021 (2001).

\bibitem{Roessl}
 E. Roessl,
 {\em Topological defects and gravity in theories with extra
      dimensions},
 arXiv:hep-th/0508099.

\bibitem{B.Bajc}
 B. Bajc and G. Gabadadze,
 {\it Phys. Lett. }  {\bf B 474}, 282 (2000).

\bibitem{I.Oda}
 I. Oda,
 {\it Phys. Lett. }  {\bf B 496}, 113 (2000).

\bibitem{Grossman}
 Y. Grossman and M. Neubert,
 {\it Phys. Lett. } {\bf B 474}, 361 (2000).

\bibitem{LiuJHEP2007}
 Y.X. Liu, L. Zhao and Y.S. Duan,
 {\it JHEP} {\bf 0704}, 097 (2007).

\bibitem{Liu0708.0065}
 Y.X. Liu, X.H. Zhang, L.D. Zhang and Y.S. Duan,
 {\it JHEP} {\bf 0802}, 067 (2008),
 arXiv:0708.0065[hep-th].

\bibitem{Jackiw1976}
 R. Jackiw and C. Rebbi,
 {\it Phys. Rev. }  {\bf D 13}, 3398 (1976).

\bibitem{Randjbar-Daemi}
 S. Randjbar-Daemi and M.E. Shaposhnikov,
 {\it Phys. Lett. }  {\bf B 492}, 361 (2000).

\bibitem{Parameswaran}
 S.L. Parameswaran, S. Randjbar-Daemi and A. Salvio,
 {\it Nucl. Phys. } {\bf B 767}, 54 (2007).

\bibitem{Libanov}
 M.V. Libanov and S.V. Troitsky,
 {\it Nucl. Phys. }  {\bf B 599}, 319 (2001).

\bibitem{Randjbar-Daemi2003}
 S. Randjbar-Daemi and M.E. Shaposhnikov,
 {\it JHEP } {\bf 0304}, 016 (2003).

\bibitem{Gregory}
 R. Gregory,
 {\it Phys. Rev. Lett. } {\bf 84}, 2564 (2000).

\bibitem{Bajc}
 B. Bajc and G. Gabadadze,
 {\it Phys. Lett. }  {\bf B 474}, 282 (2000).


 \bibitem{WangMPLA2005}
 Y.Q. Wang, T.Y. Si, Y.X. Liu and Y.S. Duan,
 {\it Mod. Phys. Lett. }  {\bf A 20}, 3045 (2005).

\bibitem{DuanMPLA2006}
 Y.S. Duan, Y.X. Liu and Y.Q. Wang,
 {\it  Mod. Phys. Lett. } {\bf A 21},  2019 (2006).







\bibitem{LiuNPB2007}
 Y.X Liu, L. Zhao, X.H. Zhang and Y.S. Duan,
 {\it Nucl. Phys. } {\bf B 785}, 234 (2007),
 arXiv:0704.2812[hep-th].

\bibitem{LiuCTP2007}
 Y.X. Liu, Y.Q. Wang and Y.S. Duan,
 {\em Fermionic zero modes in self-dual vortex background on a torus},
 {\it Commun. Theor. Phys.} {\bf 48}, 675 (2007).


\end{thebibliography}
\end{document}